\documentclass[cits]{PoS}
\usepackage{xspace}
\newcommand{\about}{$\simeq$}
\newcommand{\degree}{$^{\circ}$}

\newcommand{\Msol}{M\ensuremath{_\odot}\xspace}
\newcommand{\gam}{\ensuremath{\gamma}}
\newcommand{\flux}{ph~cm\ensuremath{^{-2}} s\ensuremath{^{-1}}\xspace}
\newcommand{\cms}{cm\ensuremath{^{-2}} s\ensuremath{^{-1}}\xspace}
\newcommand{\solar}{\ensuremath{_\odot}\xspace}
\newcommand{\Na}{\ensuremath{^{22}}Na\xspace}
\newcommand{\Al}{\ensuremath{^{26}}Al\xspace}
\newcommand{\Ti}{\ensuremath{^{44}}Ti\xspace}
\newcommand{\Sc}{\ensuremath{^{44}}Sc\xspace}
\newcommand{\Ca}{\ensuremath{^{44}}Ca\xspace}

\newcommand{\Ni}{\ensuremath{^{56}}Ni\xspace}
\newcommand{\Co}{\ensuremath{^{56}}Co\xspace}

\newcommand{\Cofs}{\ensuremath{^{57}}Co\xspace}

\newcommand{\Fe}{\ensuremath{^{60}}Fe\xspace}

\title{Gamma-Ray Line Studies of Nuclei in the Cosmos}

\ShortTitle{Cosmic Gamma-Ray Lines}

\author{\speaker{M. Leising}\thanks{Supported by NASA}\\
        Clemson University\\
        E-mail: \email{lmark@clemson.edu}}

\author{{R. Diehl}\thanks{from studies with SPI on INTEGRAL, supported by ESA and
    its member countries}\\
        Max Planck Institut f\"ur extraterrestrische Physik, Garching, Germany\\
        E-mail: \email{rod@mpe.mpg.de }}

\abstract{
Gamma-ray line studies are capable of identifying radioactive tracer isotopes
generated in cosmic nucleosynthesis events. Pioneering measurements were made 30 years ago
with HEAO-C1, detecting the first interstellar gamma-ray line from \Al, then
with SMM and numerous balloon experiments, among their results the detection of radioactivity
from supernova SN1987A, and with the Compton Observatory and its OSSE and COMPTEL
instruments in 1991-2000, which performed sky surveys in \Al and 511 keV annihilation emission and
the detection of the Cas~A supernova remnant in \Ti radioactivity. 
These measurements
have established an astronomy with \gam-ray lines, which allows us to study nucleosynthesis environments in cosmic sources. 
To date, such studies have been carried out successfully for massive stars and supernovae mainly;
radioactivities from novae and other sources are expected to exist, but have not yet been detected.
The SPI high-resolution Ge spectrometer on INTEGRAL was launched in 2002 and continues to collect data on astrophysically-important \gam-ray lines from decays of \Ti, \Al, \Fe, and positron annihilation. \Ti decay lines from Cas A have been observed with both INTEGRAL telescopes, and constrain the expansion dynamics of the ejecta. The lack of other \Ti remnants is a mystery. The \Al \gam-ray line is now measured throughout the Galaxy, tracing the kinematics of interstellar gas near massive stars, and highlighting special regions of interest therein, such as groups of massive stars in Cygnus and even more nearby regions. The detection of \Fe radioactivity lines at the level of ~15\% of the \Al flux presents a challenge both for observers and models. Positron annihilation emission from the nucleosynthesis regions within the Galactic plane appears to be mainly from \Al and other supernova radioactivity, while the bulge region's positron annihilation brightness remains puzzling.
}

\FullConference{10th Symposium on Nuclei in the Cosmos\\
		 July 27 - August 1 2008\\
		 Mackinac Island, Michigan, USA}

\begin{document}

\let\jnl=\rmfamily
\def\refe@jnl#1{{\jnl#1}}%

\newcommand\aj{\refe@jnl{AJ}}%
\newcommand\actaa{\refe@jnl{Acta Astron.}}%
\newcommand\araa{\refe@jnl{ARA\&A}}%
\newcommand\apj{\refe@jnl{ApJ}}%
\newcommand\apjl{\refe@jnl{ApJ}}%
\newcommand\apjs{\refe@jnl{ApJS}}%
\newcommand\ao{\refe@jnl{Appl.~Opt.}}%
\newcommand\apss{\refe@jnl{Ap\&SS}}%
\newcommand\aap{\refe@jnl{A\&A}}%
\newcommand\aapr{\refe@jnl{A\&A~Rev.}}%
\newcommand\aaps{\refe@jnl{A\&AS}}%
\newcommand\azh{\refe@jnl{AZh}}%
\newcommand\memras{\refe@jnl{MmRAS}}%
\newcommand\mnras{\refe@jnl{MNRAS}}%
\newcommand\na{\refe@jnl{New A}}%
\newcommand\nar{\refe@jnl{New A Rev.}}%
\newcommand\pra{\refe@jnl{Phys.~Rev.~A}}%
\newcommand\prb{\refe@jnl{Phys.~Rev.~B}}%
\newcommand\prc{\refe@jnl{Phys.~Rev.~C}}%
\newcommand\prd{\refe@jnl{Phys.~Rev.~D}}%
\newcommand\pre{\refe@jnl{Phys.~Rev.~E}}%
\newcommand\prl{\refe@jnl{Phys.~Rev.~Lett.}}%
\newcommand\pasa{\refe@jnl{PASA}}%
\newcommand\pasp{\refe@jnl{PASP}}%
\newcommand\pasj{\refe@jnl{PASJ}}%
\newcommand\skytel{\refe@jnl{S\&T}}%
\newcommand\solphys{\refe@jnl{Sol.~Phys.}}%
\newcommand\sovast{\refe@jnl{Soviet~Ast.}}%
\newcommand\ssr{\refe@jnl{Space~Sci.~Rev.}}%
\newcommand\nat{\refe@jnl{Nature}}%
\newcommand\iaucirc{\refe@jnl{IAU~Circ.}}%
\newcommand\aplett{\refe@jnl{Astrophys.~Lett.}}%
\newcommand\apspr{\refe@jnl{Astrophys.~Space~Phys.~Res.}}%
\newcommand\nphysa{\refe@jnl{Nucl.~Phys.~A}}%
\newcommand\physrep{\refe@jnl{Phys.~Rep.}}%
\newcommand\procspie{\refe@jnl{Proc.~SPIE}}%

\section{Introduction}
The potential of \gam-ray lines as a tool to study nucleosynthesis and energetic particle interactions has long been clear. Direct counting of newly synthesized radioactive nuclei as they reach \gam-ray thin regions can more clearly elucidate nuclear burning processes than any other observations, except for direct neutrino measurements. The photons are penetrating, with mean free paths of tens of g~cm$^{-2}$, so even entire galaxies are insignificant obstacles; however, thick detectors are necessary to stop them. With no \gam-ray optics available, large collection areas require large detector areas, and thus massive detection systems. Placing these above the Earth's atmosphere results in high photon and particle background rates, which reduce sensitivity and make data analysis challenging. Many predictions long preceded even the possibility of experimental verification
\cite{pmorrison1958,1965ApJ...142..189C}. A series of later predictions have guided the experimental directions of the field
\cite{1969ApJ...155...75C,1969ApJ...158L..43C,1971Natur.234..291C,1974ApJ...187L.101C,1977NYASA.302...90A,1977ApJ...213L...5R,1980ApJ...238.1017W}. Most of these measurements have now been realized.

Experimentally, the study of celestial \gam-ray lines began with Robert Haymes' Rice University high-altitude balloon flights of  NaI detectors. Pointing the wide field instrument in the direction of the galactic center, they detected a line feature near 500 keV, consistent with the electron-positron annihilation line
\cite{1972ApJ...172L...1J,1973ApJ...184..103J}. This feature had not been predicted, though in hindsight many potential sources were proposed to contribute detectable numbers of positrons. This topic is relevant to Nuclei in the Cosmos because at least some of the positrons are produced in interstellar radioactive decay. Subsequent balloon experiments confirmed this feature, including those with better energy resolution \cite{1978ApJ...225L..11L} that removed any doubt as to the origin of the line. Detection of nuclear lines from radioactive decay  awaited satellite instruments. The \gam-ray spectrometer aboard the third High-Energy Astronomy Observatory (HEAO C-1) detected the 1.809 MeV line of $^{26}$Al decay in its two two-week scans along the galactic plane in 1979 and 1980
~\cite{1982ApJ...262..742M,1984ApJ...286..578M}. That germanium spectrometer also measured the electron-positron annihilation line in the same observations~\cite{1981ApJ...248L..13R}. The Solar Maximum Mission was designed for solar flare observations, but its Gamma-Ray Spectrometer's exceptional stability and nearly a decade of tracking the Sun around sky led to improved measurements of the \Al line~\cite{1985ApJ...292L..61S,1990ApJ...362..135H} and positron annihilation line and positronium continuum~\cite{1988ApJ...326..717S}. Both the HEAO C-1 and SMM/GRS data set interesting upper limits with non-detections of line emission from \Ti and \Fe~\cite{1982ApJ...262..742M,1994ApJ...424..200L}. The unlikely event SN 1987A provided an opportunity to study a core-collapse supernova up close, and even though it could not be pointed at the supernova, the SMM/GRS was able to detect several lines of \Co decay~\cite{1988Natur.331..416M,1990ApJ...357..638L}. The lines emerged earlier than expected, indicating mixing of a few percent of the core radioactivity into the hydrogen envelope.

All of these instruments were essentially ``photon buckets'' with wide fields-of-view defined only by partially surrounding anti-coincidence shields. Long scans or large offset pointing observations for background estimation and removal were done on timescales equal to or longer than the timescales of background variations. These yielded relative flux measurements and little \gam-ray directional information. Only a tiny fraction of recorded counts were actually celestial photons. Possible systematic errors in background estimation were very hard to identify or quantify, and now seemingly erroneous reports of line detections were not uncommon~\cite{1997AIPC..410..418H}.

Advances came with the 1991 launch of the Compton Gamma Ray Observatory, one of NASA's ``great observatories.'' To study \gam-ray lines, it carried the Oriented Scintillation Spectrometer Experiment (CGRO/OSSE) and the Compton Telescope (CGRO/COMPTEL.) The CGRO/OSSE was again a photon bucket, or rather four very large ones, but it was well collimated to a $3.5^\circ \times 11^\circ$ field-of-view, and was continually offset-pointed every two minutes. This yielded unprecedented accuracy in background estimation, but produced only relative measurements of diffuse emission. The small field limited exposure to broadly diffuse emissions, and the large detector volumes brought large background rates. It produced the first maps of the diffuse electron-positron annihilation radiation and directly measured \Cofs decay in SN~1987A~\cite{1992ApJ...399L.137K}. The CGRO/COMPTEL was the first Compton telescope in space, measuring the energy loss and position in two detector planes, with the first detector optimized for a single Compton scatter. For a presumed absorption in the second detector, the photon would have to come from a ring on the sky. This ``imaging response'' allowed the discrimination of celestial photons from most background events to achieve the much improved signal to background ratio of a few percent in strong lines and direct imaging of the \gam-ray sky. The first map of the 1.809 MeV \Al decay photons was produced (Fig.~\ref{fig_26Almap_spectrum}) as well as maps of diffuse continuum emission. The detector thresholds were such that the instrument operated effectively only above 1~MeV, and some true photon background components remained difficult to identify, but the Compton telescope potential was nicely demonstrated.

The currently operating INTEGRAL instruments take a different approach, following the heritage of, e.g., the GRANAT/SIGMA experiment~\cite{1991AdSpR..11..289P}. Both the spectrometer SPI and the hard X-ray imager IBIS use coded masks in their apertures, which allows geometrical imaging and background estimation by modulating the source signals seen by different detectors at the same time. Temporal modulation by regular telescope offsets by a few degrees (``dither patterns'') offer further background reduction. For diffuse emission, the mask coding advantage is reduced, so SPI studies of, e.g., \Al and electron-positron annihilation radiation, still rely on pointing variations for modulating the signals differently from temporal changes of backgrounds. SPI employs germanium detectors, achieving an energy resolution of $E/\Delta E\simeq$500. Both of these instruments are making significant contributions to \gam-ray line studies, as outlined below. In the following sections, we discuss some of the prominent problems addressed by \gam-ray line astronomy, with emphasis on recent results.

\section{Diffuse Emission}
\subsection{$^{26}$Al}
\subsubsection{The Pre-INTEGRAL Situation}
\begin{figure}
\centering
\includegraphics[width=0.58\textwidth]{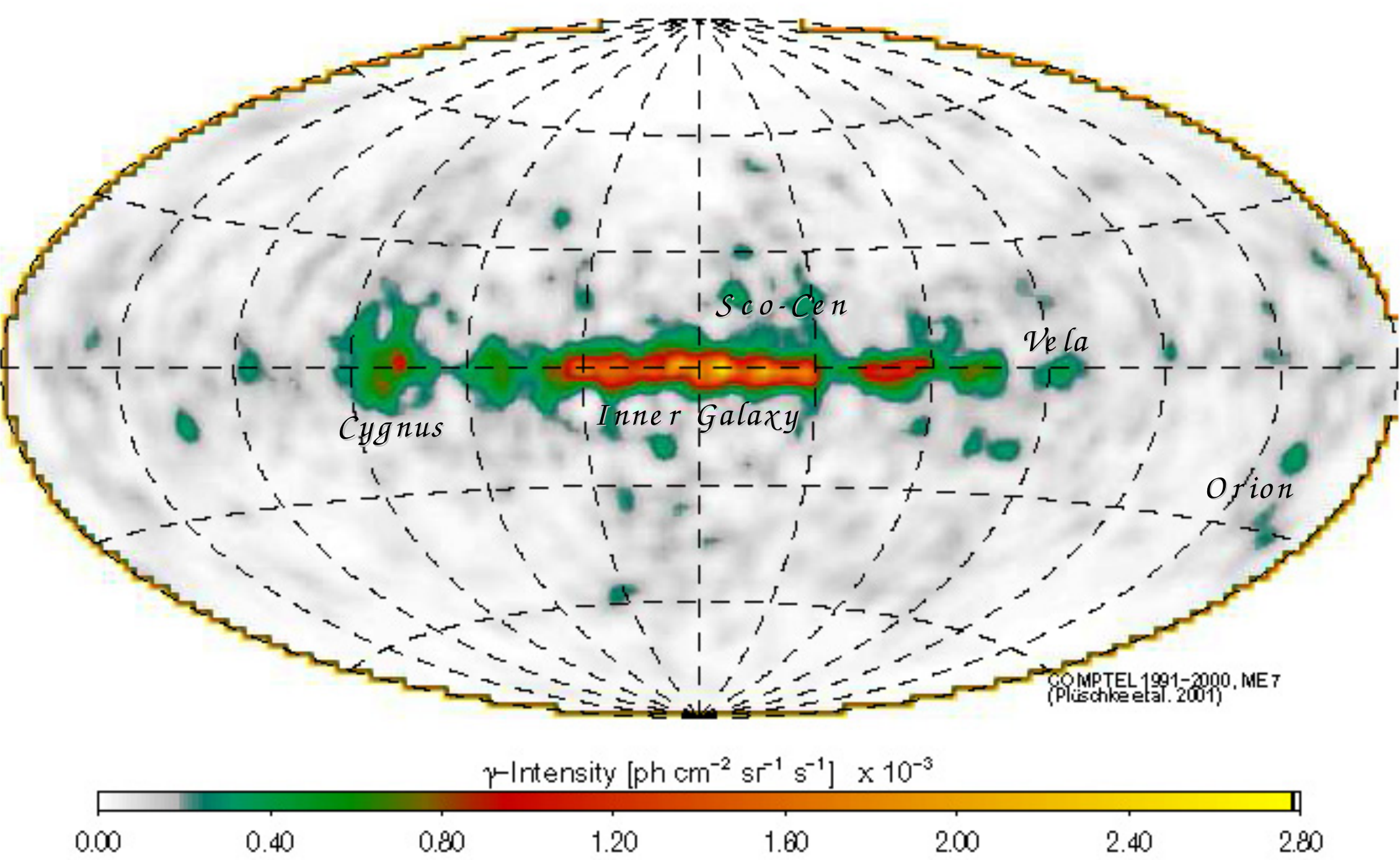}
\includegraphics[width=0.41\textwidth]{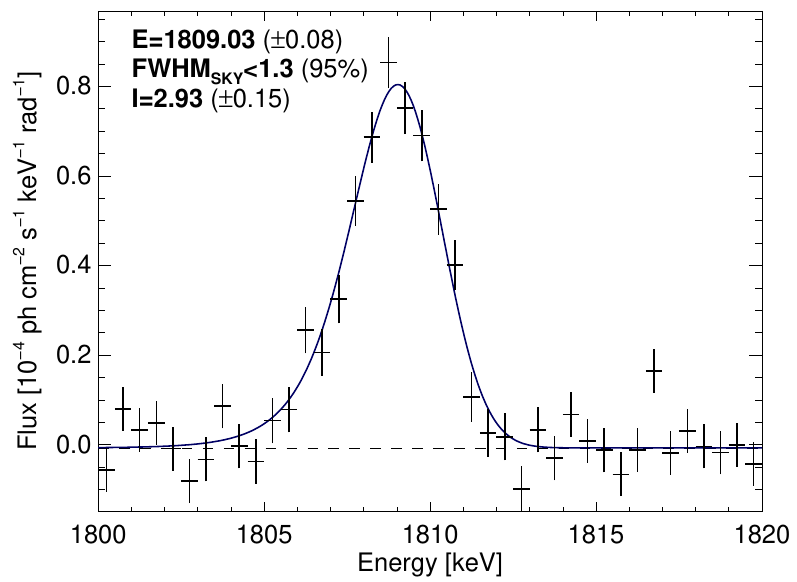}
\caption{\Al radioactivity from the Galaxy, as seen by the Compton Observatory and INTEGRAL.
    ({\it Left:}): The \Al image from 9 years of sky survey with COMPTEL shows somewhat patchy emission along the plane of the Galaxy \cite{2001ESASP.459...55P}.
    ({\it Right:}): The \Al spectrum as measured by INTEGRAL/SPI shows the line to be rather narrow \cite{2003A&A...411L.451D,2008PASP..120..118W} 
    }
\label{fig_26Almap_spectrum}
\end{figure}

The \gam-ray line discovery from \Al radioactivity ($\tau$=1.04~My, E$_{\gamma}$=1808.65~keV) with the Ge spectrometer on the HEAO-C satellite \cite{1982ApJ...262..742M} 
may be considered the most direct proof of ongoing nucleosynthesis in the current epoch of the Galaxy's evolution. The history of \Al line observations since then and up to the epoch of NASA's Compton Observatory has been described earlier \cite{1996PhR...267....1P}. 
\Al production as imaged by COMPTEL (see Fig. \ref{fig_26Almap_spectrum}~[left]) occurs throughout the Galaxy, with apparent concentrations in regions hosting young and massive stars such as in Cygnus. The patchy distribution of \Al emission along the plane of the Galaxy could partly be instrumental, yet is significant beyond this uncertainty and argues for massive stars as dominant sources of Galactic \Al. This is consistent with estimates from the ionizing radiation by those same stars \cite{1999ApJ...510..915K}. 
Adopting plausible large-scale source distributions, the total amount of \Al in the Galaxy was estimated as 2--3 \Msol. Massive stars, through core-collapse supernovae and Wolf-Rayet phases, eject \Al into the interstellar medium, where it decays to produce the observed emission. Typical ejection velocities are 1000~km~s$^{-1}$ or more. Ejecta are slowed by interaction with circumstellar and interstellar matter. Depending on when the \Al is slowed, one would expect more or less Doppler broadening of the observed \Al line. Therefore, the \Al line provides a diagnostic of the mean environment around massive stars, which is rather unique: Such interstellar gas is dilute and ionized, so radiative effects are sparse; nevertheless, this ``hot phase'' of the ISM has important implications for galactic structure and evolution. The GRIS balloone-borne Ge spectrometer instrument had reported such Doppler broadening \cite{1996Natur.384...44N}, 
but the associated mean \Al nucleus velocities at the time of decay of about 500~km~s$^{-1}$ appeared to be implausibly large \cite{1997ESASP.382..105C}. 
One of the main goals for INTEGRAL's high-resolution Ge spectrometer instrument SPI therefore was a clear spectroscopic measurement of the \Al line shape to assess the kinematics of \Al throughout the Galaxy.

\subsubsection{INTEGRAL Results}
\begin{figure}
\centering
\includegraphics[width=0.41\textwidth]{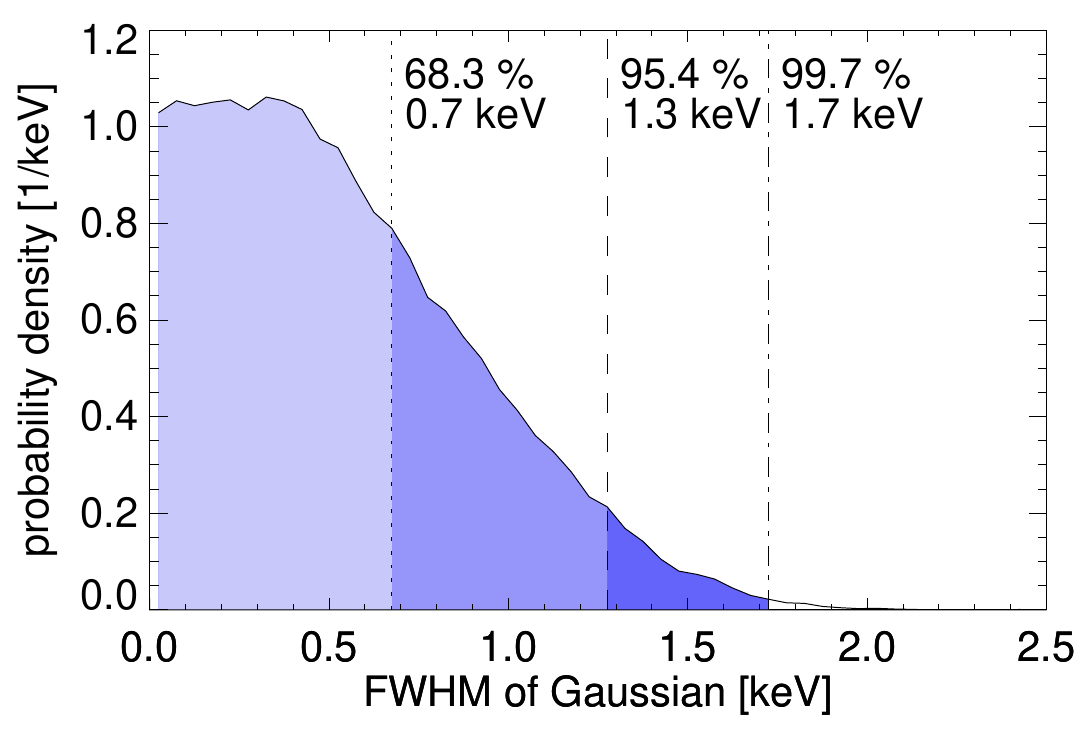} 
\includegraphics[width=0.46\textwidth]{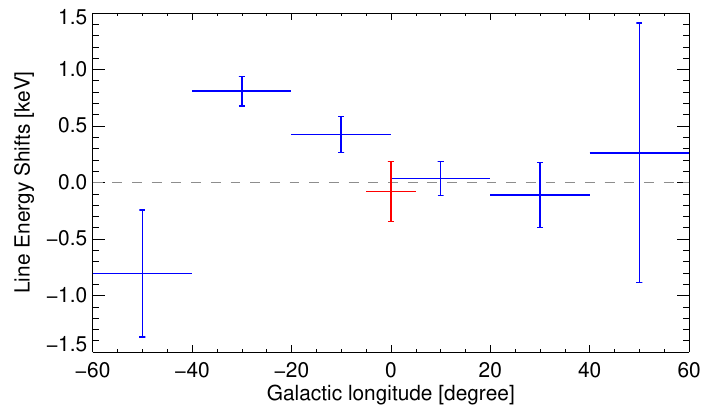}
\caption{\Al line details from INTEGRAL:
     ({\it Left:}) The probability distribution for additional broadening of the \Al line, beyond the instrumental width~\cite{2008PASP..120..118W}.
    ({\it Right:}) The \Al line centroid positions for different directions along the plane of the Galaxy show an expected blue-shift from large-scale galactic rotation in the fourth quadrant; a corresponding redshift in the first quadrant is not clear (\cite{2008PASP..120..118W}; see also \cite{2006Natur.439...45D}).}
\label{fig_26Aldetails}
\end{figure}

The \Al line was found to be rather narrow with first observations of INTEGRAL \cite{2003A&A...411L.451D}, confirming the measurement made with the Ge spectrometer on the RHESSI solar satellite \cite{2003ApJ...589L..55S}. 
Although the line was not clearly resolved, it appeared somewhat broadener than instrumental resolution, but far less than the 5.4~keV broadening of the earlier result \cite{1996Natur.384...44N}.
Subsequent INTEGRAL measurements added precision and spatial information on the \Al line parameters (Fig. \ref{fig_26Aldetails}). Observations of the inner Galaxy showed shifts in the line centroid, which plausibly followed the Doppler shifts expected from the large-scale galactic differential rotation~\cite{2006Natur.439...45D}. 
With more data, line centroid determinations can be made for different viewing directions (Fig. \ref{fig_26Aldetails}). Recent results suggest an asymmetry, with significantly-blueshifted \Al line emission towards the fourth quadrant of the Galaxy (i.e., negative longitudes). The value of the blue-shift appears on the high side of what is expected from differential rotation, and may reflect peculiar bulk motion such as expected within the bar of the Galaxy. On the other hand, there are apparently no corresponding  redshifts of the \Al line at correspondingly-positive longitude values, i.e. in the Galaxy's first quadrant. In terms of \Al emission, the inner Galaxy appears asymmetric; peculiar bulk motion could compensate for differential large-scale rotation here, or the differences in inner spiral-arm structures may be responsible for these results.

As the \Al signal weakens away from the inner Galaxy, it becomes more difficult to determine separately the line shift and width per longitude bin. Analyzing 30-degree segments, it seems that towards the Aquila direction (l$\simeq25^{\circ}$) the measured \Al line is broader than seen toward other directions. If one such region of a particularly young age (below about 5~Myrs) dominates the emission, a larger line width would be plausible, from increased turbulence of the ISM around groups of massive stars.
The integrated \Al signal observed with INTEGRAL now has reached a significance (30$\sigma$~\cite{2008PASP..120..118W}) similar to what COMPTEL obtained in its 9-year mission, making possible improved tests of kinematics of \Al nuclei at the time of their decay. The instrumental line width of 3~keV FWHM limits such tests; moreover the periodic annealings and intervening cosmic-ray induced 10\% degradations of detector resolution result in a time-variable detector resolution over the years of accumulated observations. Since these instrumental effects are well-calibrated, we can compare the observed line shape to the expected one from instrumental properties, and parametrize the difference in terms of an additional (assumed-Gaussian) astrophysical broadening. Fig~\ref{fig_26Aldetails}~(left) shows the probability distribution as obtained from this test. Clearly, there is no positive broadening measured; the probability distribution peaks at small values near zero. Yet the shape, with an indicated plateau up to a few tenths of a keV, hints towards a small broadening of 0.3--0.4 keV, which would correspond to a velocity of 50~km~s$^{-1}$. The probability distribution provides a strict upper limit of 1.3~keV (at 95\% probability) for additional broadening, which limits \Al velocities to less than 150~km~s$^{-1}$ on average. Note that this conversion to velocity limits assumes an isotropic and Gaussian distribution of velocities, such as expected from turbulence in the interstellar medium. Also, the underlying spectrum is obtained from integration over the entire inner Galaxy, which may include regions of different intrinsic turbulence plus the bulk motion from the Galaxy's differential large-scale rotation. In summary, large kinematic broadening in the 100~km~s$^{-1}$ range such as discussed ten years ago are excluded, while INTEGRAL's capability to measure velocities down almost to average stellar velocities provides an interesting and unexpected perspective.

\subsubsection{Issues for Future Measurements}
Clearly, the integration over many possibly unrelated source regions limits the impact of \Al results obtained so far. Only in the Cygnus region may one safely assume a coherent and localized single source region along the line of sight, which also has a strong \Al signal, on the order of 15$\sigma$ \cite{2004ESASP.552...33K}. However, the INTEGRAL exposure of the inner Galaxy now is approximately similar to the COMPTEL all-sky survey (1991--2000). With similar spatial resolutions of INTEGRAL/SPI and COMPTEL (2.7\degree versus 3.6$^{\circ}$ FWHM, respectively), this allows study of specific regions of interest for local line shape diagnostics, with years of INTEGRAL observations still to come. The peculiar line width towards the Aquila region, and isolation of the \Al signal from the nearby Sco-Cen association (distance 100--150~pc) are on the horizon. It will be important that such observations are taken under well-controlled background and spectral-response conditions, so that instrumental limitations are minimized. The Orion region presents the most nearby concentration of young massive stars at about 450~pc distance, well isolated on the sky with respect to other Galactic \Al sources, and marginally detected with COMPTEL (Fig.~\ref{fig_26Almap_spectrum}) \cite{2002NewAR..46..547D}. Similar to Cygnus, or even better, is our knowledge about the stellar census of this region. But, being located towards the outer Galaxy with correspondingly lower space density of candidate hard X-ray sources, INTEGRAL's observing program devoted just as much exposure to Orion so that confirmation or not of the weak COMPTEL \Al signal may be expected. These data are being analyzed, and a positive \Al detection would likely prompt more observations, for a determination of bulk \Al velocity of Orion. This is of interest because there is a large interstellar cavity in the foreground of the Orion \Al sources, which plausibly could cause deviations from spherically-symmetric \Al dispersion around its sources.

\subsection{$^{60}$Fe}
Stellar nucleosynthesis models have shown that the long-lived \gam-ray emitting radioactive isotope \Fe plausibly is synthesized within the convective envelopes of massive-stars. For stars exceding 8--10 \Msol, these \Fe-rich ashes would be ejected in the terminal core-collapse supernova explosion. The details are complex, and how shell burning regions develop and extinguish and which neutron-producing reactions are important are uncertain. Yields of radioactive \Al and \Fe are very sensitive to such detail, however \cite{2002NewAR..46..459C}. Because of the dominant massive-star origin of \Al, the \gam-ray flux ratio from the two isotopes constitutes an interesting global test of massive-star nucleosynthesis models.

Following many reported non-detections and upper limits, a hint of \Fe \gam-rays was found in 2003 with RHESSI \cite{2004ESASP.552...45S}. 
The weak signal corresponded to 16$\pm$5\% of the \Al flux,  consistent with earlier theoretical predictions of ~15\% \cite{1996ApJ...464..332T}. 
Later studies of massive-star nucleosynthesis tended to predict larger ratios of \Fe versus \Al, as progenitor evolution and wind models\cite{2004A&A...420.1033P}, as well as nuclear-reaction rates\cite{2007PhR...442..269W}, were updated . 
Nucleosynthesis calculations \cite{2002NewAR..46..459C,2007PhR...442..269W} generally still fall on the higher side of the original prediction, but are consistent, given the substantial uncertainties in such models. Uncertainties arise mainly from stellar structure, as establishment of suitable convective-burning regions is sensitive to stellar rotation, which in turn is affected by the mass loss history during evolution. Uncertainties on nuclear cross sections involve \Al destruction though n~capture, and n-capture on unstable $^{59}$Fe and on \Fe itself.
\begin{figure}
\centering
\includegraphics[width=0.37\textwidth]{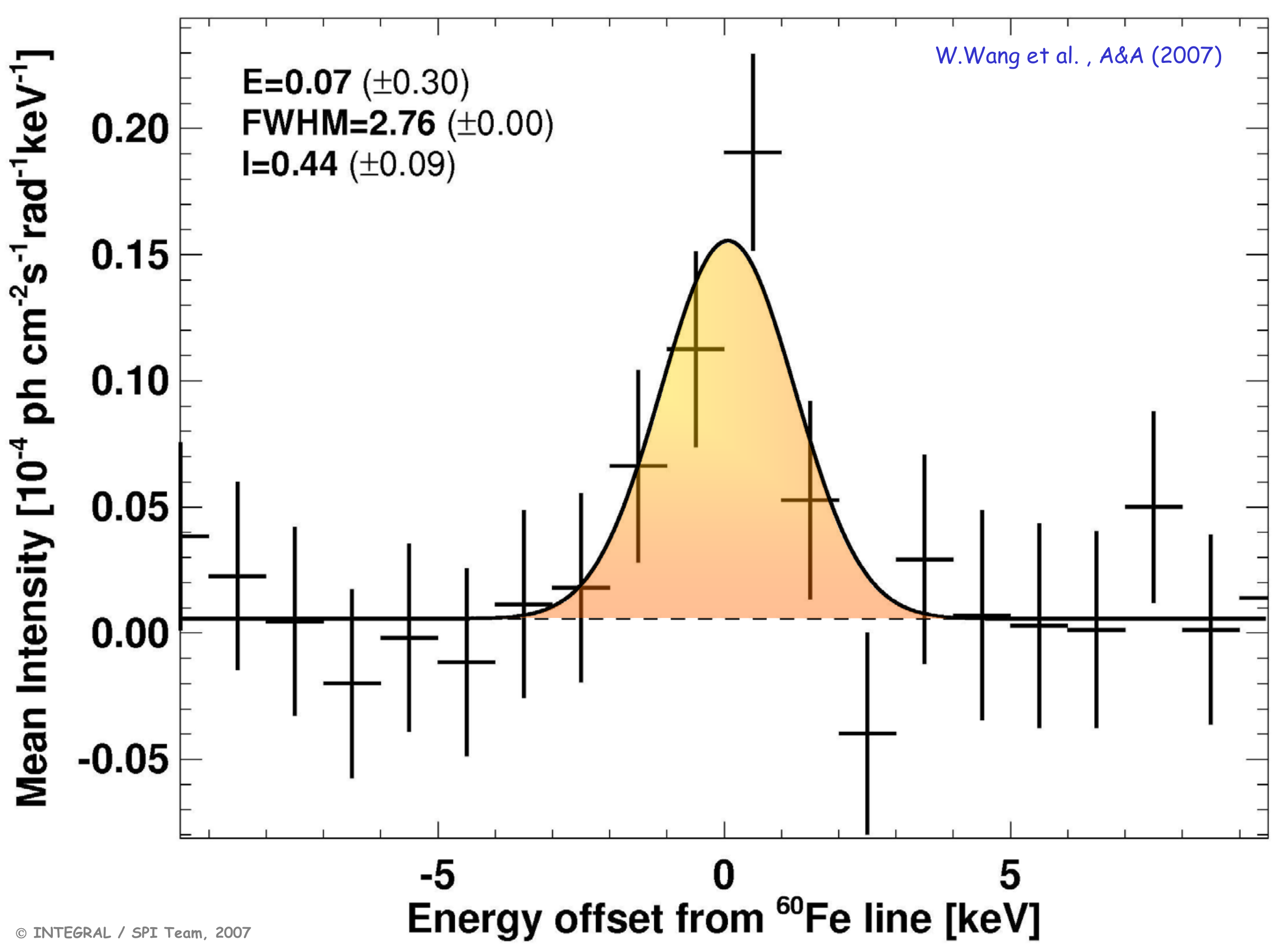}
\includegraphics[width=0.58\textwidth]{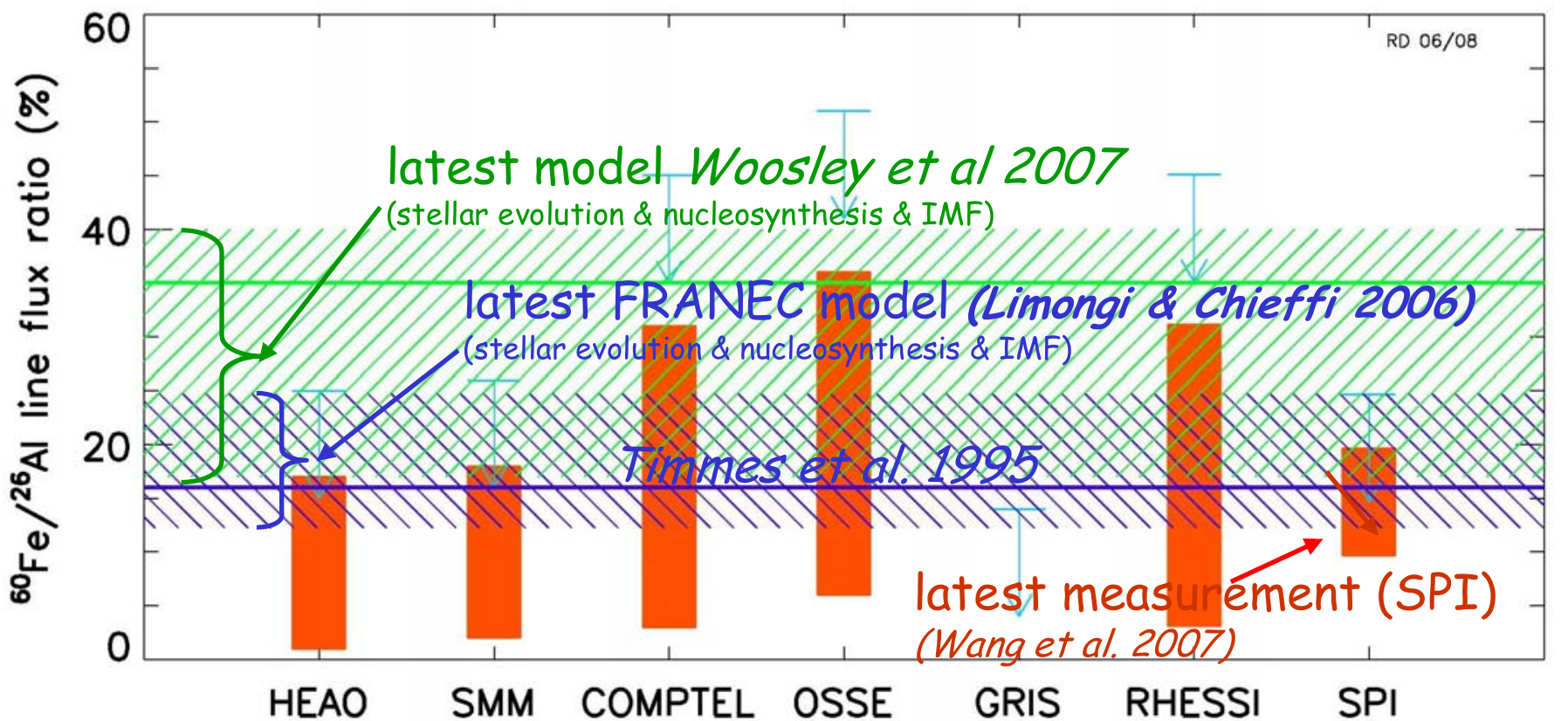}
\caption{
     ({\it Left:}) The \Fe measurement with INTEGRAL . Superpositions of both lines of \Fe at their nominal energies of 1173 and 1332~keV, respectively, show a weak signal from \Fe decay.
    ({\it Right:}) The set of \Fe/\Al ratio measurements by different experiments and the ranges of current theoretical predictions appear consistent. Deeper investigations of uncertainties in measurement and model may show if observations are significantly lower than predictions.}
\label{fig_60Fe26AlRatios}
\end{figure}

Confirmation of the RHESSI \Fe signal was reported from first-year INTEGRAL/SPI data \cite{2005A&A...433L..49H}, 
although features from a nearby instrumental line indicated systematics issues. In a recent analysis of more data, a significant \Fe signal (at 5$\sigma$) was found, with somewhat reduced systematic effects from instrumental background. This underlying background is being investigated, specific signatures within the 19-detector Ge camera of SPI are being exploited to discriminate internal versus celestial \gam-rays on their respective modulation time scales. The INTEGRAL/SPI reported \Fe/\Al \gam-ray flux ratio is now 0.14$\pm$0.06. 
Formally, there is agreement between observations and models (see Fig. \ref{fig_60Fe26AlRatios}), but more can be learned as uncertainties in each area are revisited and re-assessed. We also hope to exploit INTEGRAL's spatial resolution, towards determination of spatially-resolved \Fe to \Al ratios. It seems feasible to obtain values for the different Galactic quadrants, and the \Fe limit for the \Al-bright Cygnus region will provide another interesting constraint because, rather than being in steady state, here \Al from rather young massive-star groups is observed.

\subsection{Electron-Positron Annihilation Radiation}
As described above, this subject has been studied experimentally for nearly four decades. We still do not understand the origin of the positrons, though there are many possible sources. After the initial discovery~\cite{1973ApJ...184..103J} numerous balloon-borne instruments apparently measured different line fluxes, in some cases just months apart. The common interpretation was that a necessarily compact source, from light-travel time arguments, was episodically ejecting positrons into a relatively dense medium where they could annihilate quickly, i.e., the "Great Annihilator"~\cite{1991SciAm.265...29L}. The largest line flux was measured by the SMM/GRS, and was found to be constant over several years~\cite{1990ApJ...358L..45S}. Many of the earlier differing fluxes could be explained by the different fields-of-view of the various instruments viewing a broadly diffuse distribution of annihilation sites~\cite{1991AdSpR..11..203V}.

This diffuse distribution was naturally thought to correspond to the galactic disk, because so many of the most plausible positron sources were located there. These include supernova produced radioactive positron emitters, especially \Al and \Sc (from \Ti) from core collapses and \Co (from \Ni) from thermonuclear events, pairs produced in black hole high temperature accretion disks or jets, pulsar pair winds, and cosmic ray/ISM collision produced positron emitters or $\pi^+$'s, among others. Thus it was somewhat surprising that early CGRO/OSSE measurements showed that the highest flux measured in its small field came from near the galactic center and was nearly independent of the direction of the rectangular collimator~\cite{1993ApJ...413L..85P}. The sky distribution of the emission was fit with a compact bulge-like component, and a more extended disk.

As more observations accumulated at a variety of positions and collimator angles, it became possible to  find model-indpendent source distributions roughly consistent with the data~\cite{1997ApJ...481L..43C,1997ApJ...491..725P}. These showed similar features: a compact bulge with an intensity profile with FWHM $\simeq 5^\circ$, an extended component along the Milky Way disk, and an excess slightly offset from the galactic center to negative longitude and positive latitude. This third component was detected with relatively low ($\simeq$3.5$\sigma$) statistical significance, and much attention followed it, but the origin of the brightest component remained a mystery. More data were acquired, and statistics in both mapping and model fitting improved~\cite{2001ApJ...559..282K,2001AIPC..587...11M}. The significance of the feature at positive latitude diminished. The basic description of the emission remained a bright central bulge and a relatively weak disk. The line and positronium continuum fluxes  were everywhere consistent with annihilation of 95\% of the positrons via the formation of positronium in the statistical ratio of 3:1 for the triplet to singlet states; the triplet state annihilation continuum was detected with greater significance than the line. Because these remained relative flux measurements due to the offset pointing background estimation,  the absolute fluxes of bulge and disk components were not tightly constrained. The central component could be compact, like the stellar bulge, with a relatively low total flux, or more extended, like the halo stellar surface brightness, with a larger total flux. Similarly, a very thin disk with a low flux fit nearly equally as well as a vertically extended disk with much larger total flux. The ratio of bulge flux to disk flux, and by assumption, positron production rates, was found to be of order unity, with an uncertainty almost a factor of three. The flux of each component was about $7\ 10^{-4}$ \cms, with a systematic uncertainly in each of roughly a factor of two. Note that both components cannot be pushed to their lower limit together, which would be inconsistent with the wide-field measurements, that of SMM/GRS in particular.

Any such bright bulge was unexpected, as most of the prominent sources are in the young stellar population of the disk. Even older population sources, such as Type Ia supernovae, would be expected to inject substantially more positrons into the massive disk than the bulge. It seems that multiple sources are required, and known or very likely sources such as the decay of \Al and \Ti should be subtracted from the disk, raising the bulge to disk flux ratio for the unknown source(s). The annihilation line flux from \Al positrons is F$_{511}$ =  0.47~F$_{1809}$ for the above positronium fraction, so the 1.809 MeV flux above indicates that 0.1 to 0.5 of the disk annihilation flux is due to \Al decay. The \Ti decay chain is not so directly observed, but various arguments (outlined below)  suggest that the average rate of \Ti production provides a line flux at 511 keV of $2\ 10^{-4}$ \cms, coincidentally similar to that from \Al. Still it seemed that an additional source of positrons was required in the disk, with a still more productive source in the central bulge or halo.
\begin{figure}[th]
\centering
\begin{minipage}[c]{0.44\textwidth}
  \includegraphics[width=0.99\textwidth]{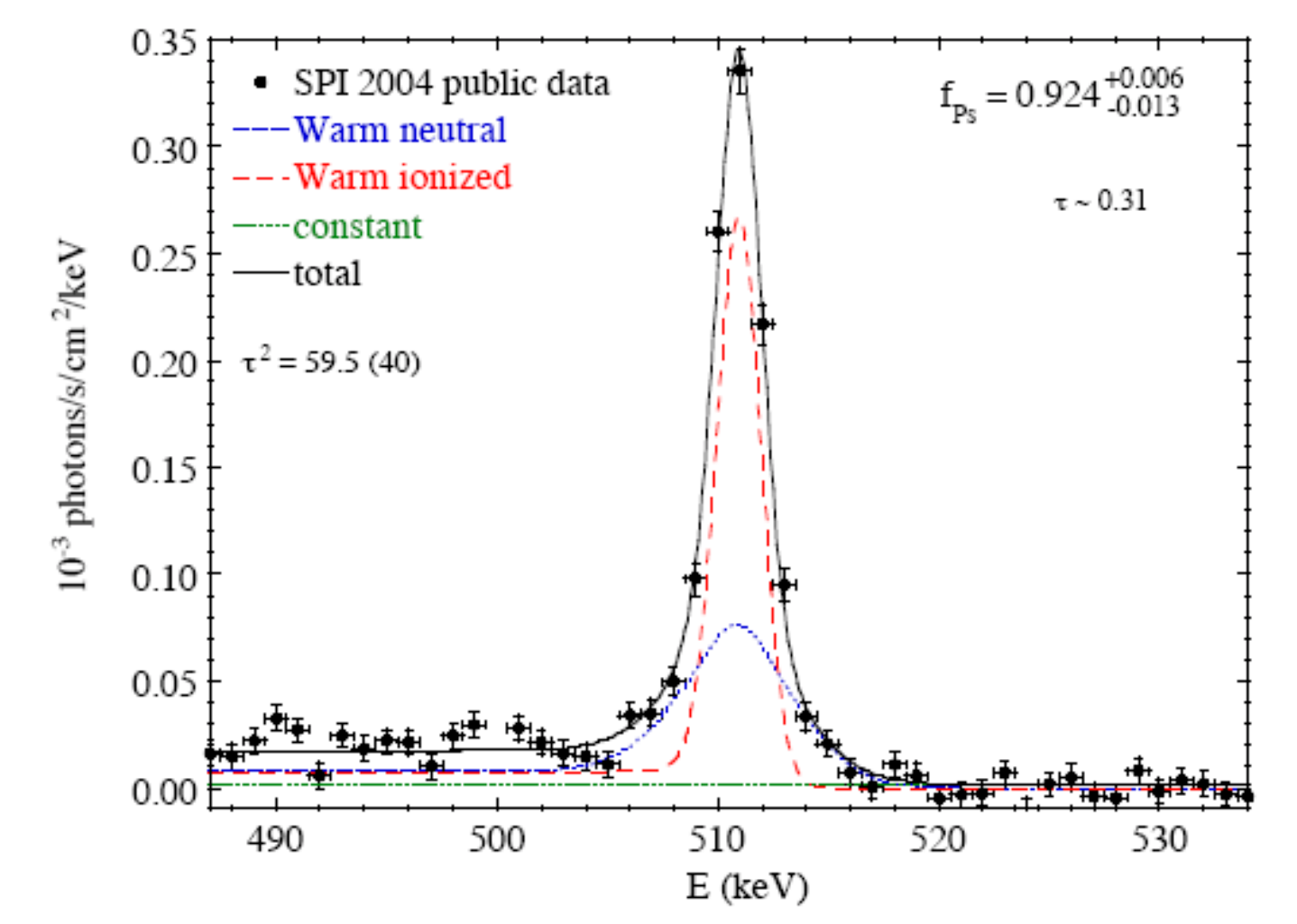} 
\end{minipage}
 \begin{minipage}[c]{0.54\textwidth}
  \includegraphics[width=0.99\textwidth]{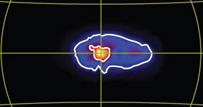}
  \end{minipage}
\caption{
({\it Left:}): The annihilation line spectrum as measured by INTEGRAL/SPI~\cite{2005MNRAS.357.1377C}.
    ({\it Right:}): The intensity of the annihilation line between galactic longitudes $\pm 60$\degree and latitudes $\pm 30$\degree, derived by an iterative maximum likelihood method (adapted from~\cite{2008Natur.451..159W}.)
}
\label{fig_SPIann}
\end{figure}

INTEGRAL/SPI has rediscovered many of these aspects of the electron-positron annihilation radiation. The SPI energy resolution is a major improvement over previous instruments with comparable sensitivity. Its relatively large field-of-view accumulates exposure to the diffuse emission efficiently, especially given the INTEGRAL mission's emphasis on the inner Galaxy. Analyses of the early SPI data showed the central bulge-like component dominating the emission, with a flux similar to the CGRO/OSSE value~\cite{2005A&A...441..513K} and a slightly larger extent, 8$^\circ$ FWHM if fit with a two-dimensional angular gaussian.
The disk component was not yet clearly detected. The annihilation line was resolved, and the line profile suggested two distinct components were present: a broad component from annihilation in  a warm (several thousand Kelvin) neutral medium in which positronium formation by charge exchange with H and He dominates, and  a narrow component from a warm ionized medium in which coulomb losses drive the positrons below the charge exchange energy threshold and positronium forms radiatively~\cite{2005MNRAS.357.1377C,2006A&A...445..579J}. The positronium triplet state continuum distribution was found to be similar to the line's, with a dominant bulge and relatively weak disk~\cite{2006A&A...450.1013W}. All of these studies used only a small fraction of the data now available.

Recently, analyses of more than three times as much SPI data (5.4$\ 10^7$ s) have been performed~\cite{2008Natur.451..159W}. Similar bulge fluxes (~$10^{-3}$ \cms) and extent (6$^\circ$ FWHM) are found, still subject to the same systematic uncertainties. The disk is now clearly detected, and evidence is presented that the disk is asymmetric, with the negative longitude inner disk brighter than the positive longitude side. The disk is fit with a broader (7$^\circ$) latitude profile, so the total flux is comparable to that of the bulge, and the negative longitude side is brighter by a factor 1.8. It remains true for these SPI data that the measured fluxes are model dependent for each component. Lower flux is obtained for a more compact bulge, here fit with overlaid azimuthally symetric gaussian functions with FWHM of 3.4$^\circ$ and 11.6$^\circ$, than a more extended halo-like function. Similarly, the latitude extent of the disk emission is poorly constrained; wider disks have significantly higher fluxes.

An independent analysis of nearly the same data set using a different background estimate finds a quite similar bulge extent and flux~\cite{2008ApJ...679.1315B}. The disk is fit with a large flux, 1.7$\ 10^{-3}$ \cms, and is found to be symmetric in longitude and have a latitude extent 15$^{\circ}$--20$^{\circ}$. Clearly we are still learning how to best analyze these data, and substantially more data are newly obtained or soon will be, so we can expect further clarification and possible surprises.

Progress is slow in understanding the origin of the positrons, and indeed, how many are due to nuclear processes. The possible disk asymmetry has been noted to be similar to the asymmetric distribution of hard sources among low-mass X-ray binaries~\cite{2008Natur.451..159W}. These sources are also found in the galactic bulge, so they could possibly explain that positron component, but it seems the ratio of bulge positrons to disk positrons, after subtraction of the \Al and \Ti contributions, is still larger than the ratio of numbers of hard LMXB sources. However, even if this correlation proves to be true, we do not know whether these hard sources eject the positrons themselves, or whether they might be tracers of another population that does. We also note that the \Al longitude profile shows an asymmetry in longitude in the same sense as the annihilation line flux~\cite[and Fig.~\ref{fig_26Almap_spectrum}]{2000AIPC..510...35P}. Again, the direct flux from \Al decay positrons is insufficient to explain the entire positron disk or its asymmetry, but it might point to a young stellar population source. Another plausible source of positrons in the old bulge is \Co produced in Type Ia supernovae. Only a few percent of the of positrons produced at the expected rate need to escape the ejecta to contribute significantly to the observed signal. This is not implausible, but the expected rate of SN~Ia in the disk is thought to be significantly larger than that of the bulge, so if the SNe are not intrinsically different, the relative positron annihilation rates in the two components rule out these as the dominant source. One of the few bulge-only sources, low-mass dark matter particles that decay or annihilate into pairs~\cite{PhysRevLett.92.101301} is an interesting possibility, but not yet constrained by the \gam-ray data.

\section{Specific Sources}

\subsection{$^{44}$Ti Supernova Remnants}
Given a galactic supernova rate of 2--3 per century and a \Ti ($^{44}Ti\ \frac{\tau=87\,y}{68,78 \,keV}\ ^{44}Sc \ \frac{\tau=5.7\,h}{1156\,keV}\ ^{44}Ca$) yield of nearly 10$^{-4}$ M\solar per event, it is reasonable to expect to find a few remnants in the inner Galaxy detectable at line flux levels near 10$^{-4}$~\cms. Increasingly sensitive searches found none there~\cite{1992ApJ...387..314M,1994ApJ...424..200L}, but the Cas~A supernova remnant  was detected in the 1.16 MeV line~\cite{1994A&A...284L...1I}. This has been confirmed by detections of the lower energy lines~\cite{2001ApJ...560L..79V,2006ApJ...647L..41R}, and now all flux measurements are consistent with $\simeq$2.5$\ 10^{-5}$~\cms. These verify the idea that \Ti is synthesized in the alpha-rich freezeout of nuclear statistical equilibrium, which does occur in material ejected by core-collapse supernovae. Such observations might shed light on the role of asymmetries in the explosion and on the mechanism of ejection. INTEGRAL/SPI has begun to do spectroscopy on the Cas~A \Ti lines, with a chance to make use of simultaneous measurements of all three decay lines with the same instrument, thus minimizing systematics (see below).

The lack of other detectable remnants is quite puzzling viewed in any of several ways. Star formation is concentrated in the inner Galaxy, and current \gam-ray line surveys~\cite{2004ESASP.552...81R} can detect typical remnants at the distance of the galactic center for a few half-lives of \Ti, so the probability that the brightest remnant is in the outer Galaxy, at a distance of 3.4~kpc and of an age 320 years is very improbable, assuming Cas~A's \Ti yield is not much different than the average. Apart from Cas~A, current estimates of supernova rates and yields suggest that at least a few remnants should be detected in the inner Galaxy~\cite{2006A&A...450.1037T}, but even those rates and yields fall short of the expected \Ca production rates required to explain the solar \Ca abundance~\cite{1994ApJ...424..200L,1996ApJ...464..332T}. A scenario that reproduces the solar abundance of \Ca, which we think must be made as \Ti, yields even more detectable remnants.

The solution of this puzzle is not clear. Perhaps supernova \Ti yields vary greatly, and the outliers on the high side contribute most of the \Ti. However, Cas~A, whose yield, 1.4$\ 10^{-4}$ M\solar is higher than theoretical calculations, is not extreme enough. We would still require 1--2 such supernovae per century, a few of which should be detectable. A plausible solution is that much higher yield \Ti sources, which are very rare, e.g., He-triggered low-mass thermonuclear supernovae~\cite{1986ApJ...301..601W}, produce the time-averaged \Ca rate, but none has occurred in the Galaxay in recent centuries. No such source has been recognized among extragalactic supernovae or galactic remnants. Such objects should be visible in the K X-rays of radioactive $^{59}$Ni ($\tau=$75 ky) decay~\cite{2001ApJ...563..185L} if they exist.

There is also evidence that \Ti is present in the ejecta of SN~1987A~\cite{2001A&A...374..629L} at a mass somewhat less than that in Cas~A. The implied \gam-ray line flux could be 3$\ 10^{-6}$~\flux, which is undetectable to current instruments, but could be measured by near-term instruments with hard X-ray optics or much improved Compton telescopes.

\subsection{Supernovae Ia and Novae}
From simple considerations, it was thought that the objects best studied by \gam-ray lines would be the profoundly thermonuclear objects, classical novae and Type Ia supernovae, rather than the core-collapse supernovae for which the nuclear reactions are a side effect. However, we have definitively detected neither a nova nor a thermonuclear supernova. The reason is simply luck; none have occurred within the suitable volumes sampled by our instruments.

Type Ia supernovae are good standard candles because they are efficient C and O thermonuclear bombs, producing typically 0.6~M\solar of \Ni whose decay powers the visible display. Their kinetic energies are high enough that the \gam-ray escape is significant after one month. The best limit on \Co lines from a SN~Ia was set by the SMM/GRS for the nearby SN~1986G, which found to have ejected less than 0.4~M\solar of \Ni~\cite{1990ApJ...362..235M}. This was a somewhat under-luminous Type Ia, for which that \Ni mass is not now thought to be extreme. The high-luminosity SN~Ia prototype, SN~1991T, was the first prospect for CGRO, but its distance of 13--14 Mpc meant that only upper limits were achieved~\cite{1994A&A...292..569L,1995ApJ...450..805L}, however one analysis method showed possible \Co line features~\cite{1997AIPC..410.1084M}. SN~1998bu was only slightly closer, and also yielded only upper limits~\cite{2002A&A...394..517G}. No SN Ia has been closer in the INTEGRAL era, though SN~2003gs was observed early in the mission. No lines were seen.

Classical novae, envelope thermonuclear flashes on accreting white dwarfs, were long suspected to be  bright \gam-ray line sources \cite{1974ApJ...187L.101C,1987ApJ...323..159L} Proton rich nuclei, including \gam-ray emitters \Na and \Al and short-lived positron emitters $^{13}$N and $^{18}$F, should be ejected, but yields are uncertain mainly because the underlying models are uncertain. Searches for \Na lines have so far proved fruitless, both blind and of nearby novae~\cite{1982ApJ...262..742M,1988ApJ...328..755L,1995A&A...300..422I}.   The nova contribution to the diffuse \Al emission is unknown, but is probably not dominant given the lack of \Al in the bulge. The electron-positron annihilation emission over the first hours after outburst would be an excellent diagnostic of the burning and transport of envelope material. The models have become somewhat more pessimistic in recent years~\cite{1999ApJ...526L..97H}. As the onset of the runaway is unpredictable, a wide field monitor with good background stability is essential. The FERMI \gam-ray burst monitor (GBM) might serve this purpose in the immediate future.

\section{Science Issues for the Future}
\subsection{Thermonuclear SNe}
Though we have had the basic model of Type~Ia supernovae for nearly twenty-five years, many of the details still elude us [Woosley, this volume.] What the progenitor system is, how and where the runaway ignites, how the flame propagates, and how these factors affect the observed correlation between peak brightness and decline rate are not yet clear. Some of these could begin to be deduced from a detailed measurement of the run of the \Ni abundance within the ejecta. Gamma-ray line astronomy can approach this in two ways, by measuring the line profiles as they emerge, which are determined by a combination of Doppler and attenuation effects, or by measuring the evolution of the line fluxes over several months, which are determined by how the total overlying eject thin relative to the \Ni location. The expansion speeds of $\simeq$10,000~km~s$^{-1}$ give line widths of a few percent. Energy resolution better than 100 is therefore required for the former objective. Ge-detectors, for example, which gain sensitivity for narrow lines by spreading the background away from the line energies, lose that advantage for broad lines. For this reason, INTEGRAL/SPI is sensitive to SNe~Ia to distances of $\simeq$6~Mpc. 

For a comprehensive attack on this problem, much improved sensitivity is paramount -- to detect the lines with high signal-to-noise, and to detect a number of SNe within each of the ranges of peak visible luminosity now recognized. A broad line sensitivity of 1$\ 10^{-6}$~\flux would allow us to detect SNe~Ia to $\simeq$60 Mpc, but more importantly, we would measure one per year within 20~Mpc at significance $\geq$30$\sigma$ in each of multiple lines at several epochs. This would revolutionize our understanding of the SNe~Ia. 

\subsection{Core collapses and GRBs}


Core-collapse supernovae are at the heart of many of the evolutionary processes in the universe. They are the end points of stellar evolution for stars with masses in excess of \about 8 \Msol, and the birth sites of neutron stars and black holes. The most-luminous single sources in the entire universe, \gam-ray bursts (GRBs), are most likely related to these outcomes of massive-star evolution. The long-duration GRBs are currently understood to arise from catastrophic collapse of massive stars under the influence of stellar rotation: angular-momentum transport leads to strongly non-spherical collapse, forming an accretion disk around a newly-forming black hole, whereby polar regions are sufficiently diluted to allow the escape of a violent jet of electromagnetic energy.

Typically, several \Msol of envelope material hide all this from the outside observer. It is difficult to study the detailed processes in the very interior of a collapsing star, which determine the outcome of stellar collapse. Direct studies probably must be deferred to astronomical observations with neutrinos and/or gravitational waves. Indirect studies are what most of our present understanding relies on.
Following a core collapse, a small fraction of the neutron star binding energy is transformed into the internal energy of the star (about 10$^{51}$~ergs); burning of all material to nuclear statistical equilibrium occurs to a radius of \about4000~km.
Radioactive \Ni is dominant,  and \Ti increases inward with peak temperature. Lower density at freezeout temperature, which favors more free $\alpha$-particles and more \Ti, occurs for matter expanding from initially higher temperature. How much of each isotope is ejected depends on pre-supernova stellar structure and the details of the ejection dynamics, which are poorly understood. Their abundances in the ejecta are excellent diagnostics.  In particular, the relative ratios of these isotopes encodes the bifurcation between ejected and collapsing inner material (the ``mass cut''). Studying how this quantity varies among supernovae, and with their mass, stellar rotation, metallicity and progenitor evolution might clarify whether there is a continuum of phenomena to the extremes of such collapses that produce black holes and \gam-ray bursts. 
\begin{figure}
\centering
\includegraphics[width=0.44\textwidth]{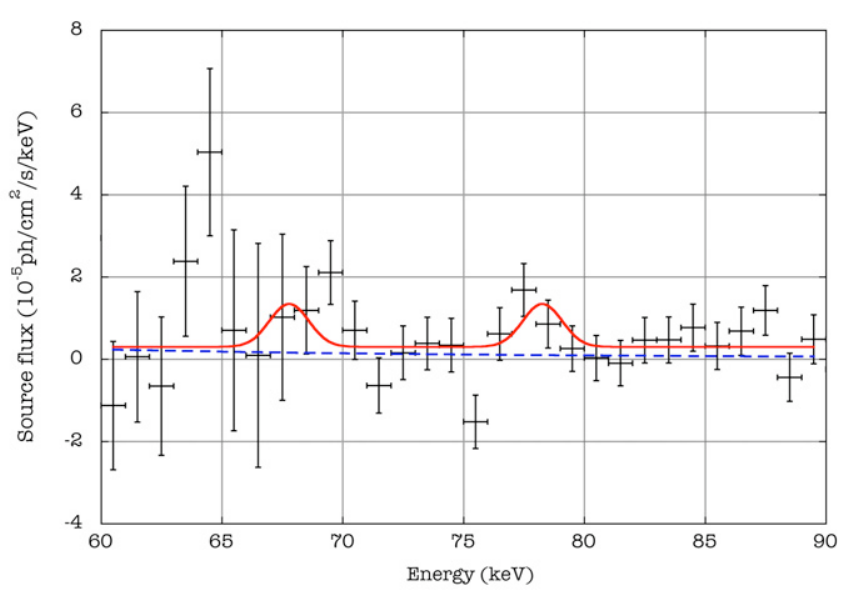}\hfil
\includegraphics[width=0.46\textwidth]{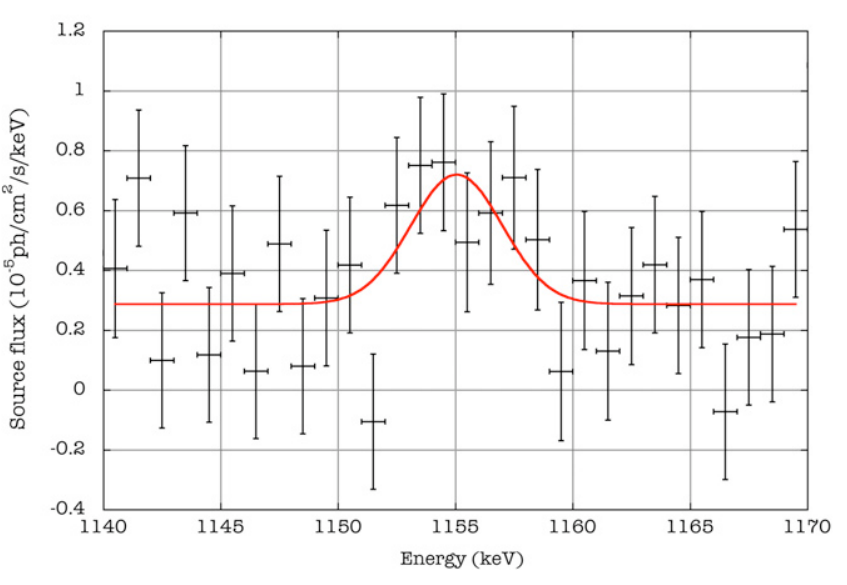}
\caption{INTEGRAL/SPI measurements for the different lines associated with \Ti decay at 68, 78, and 1156~keV. The line detections still are marginal, yet illustrate the potential to constrain energy-dependent Doppler broadening through a single-instrument spectrum, as intensities of all three lines must be identical (adapted from \cite{2008NewAR..52..401M}.) }
\label{fig_44Ti_CasA}
\end{figure}

A first diagnostic is the spatial distribution of those radioactivities in the reference frame of the collapse, and their ejection velocities. SPI is beginning to simultaneously constrain the high-energy line at 1156~keV with at least one of the low-energy lines at 78~keV (Fig. \ref{fig_44Ti_CasA}). While the low-energy lines provide a rather reliable measurement of the total \Ti \gam-ray intensity, the Doppler broadening (which is linear in energy) may broaden the 1156~keV line well beyond instrumental line widths such that it eventually drowns in the continuum background. From first studies, the fact that a signal emerges for the high-energy line determines a limit on such Doppler broadening, which translates into a velocity spread for ejected \Ti of 430~km~s$^{-1}$ at an integrated bulk velocity (from the line shift) of 500~km~s$^{-1}$, both values being quite uncertain from the still-marginal line signals (Fig. \ref{fig_44Ti_CasA}) \cite{2008NewAR..52..401M}. With dedicated deep observations, INTEGRAL has a realistic chance to obtain a significant velocity determination. Combined with fine \Ti imaging at the arcsec level, which is on the horizon with the NuStar \cite{2005ExA....20..131H} 
and Simbol-X \cite{2008MmSAI..79...19F,2008MmSAI..79...50R} 
missions, interesting constraints may be placed on core-collapse dynamics.


The immediate environment of a violent supernova explosion is rapidly ionized. Few diagnostic transitions are accessible in  such ionized material. For the case of \gam-ray bursts, there is, however, an interesting chance of still detecting such highly-diagnostic circum-burst material through {\it nuclear} transitions. Nuclear absorption may be expected from foreground material illuminated by the GRB and afterglow \gam-rays, through nuclear-level transitions in the MeV range, and giant resonance (near 25~MeV) or nucleonic delta resonance (near 325~MeV) absorptions. The strongly-beamed GRB emission acts as an amplifier over continuum absorption processes for the expected resonance absorption features. Estimates show that plausible material column densities around GRBs range up to 10$^{29}$~cm$^{-2}$, while column densities of 10$^{25}$~cm$^{-2}$ and beyond could be detected with currently-proposed instrument designs \cite{2008ExA...tmp...23G}. 
This provides another interesting perspective on nuclear processes as new astronomical tools for the most energetic stellar processes in the universe.


\subsection{Future Instruments}
In comparison to adjacent astronomical bands in the electromagnetic spectrum from radio to TeV $\gamma$-rays, the band of nuclear astronomy lags behind in sensitivity by a few orders of magnitude -- only the brightest and most-nearby sources have been seen up to now. Technological efforts focus on an extension of focussing optics as far as possible into the range of penetrating \gam-rays, and on interaction-tracking detector systems with highest resolutions in temporal and spatial parameters (see~\cite{2004NewAR..48....1D} for a review of instruments).

\subsubsection{Hard X-ray Mirrors}
Improved sensitivity to \gam-ray lines is essential for future instruments. The conventional astronomer's approach, a large-area collector focusing on a small volume detector, can at least be employed with grazing-incidence mirrors at hard X-ray energies. The main limitation is the relative low energies that can be reflected. Some progress seems assured. The NASA Small Explorer NuSTAR~\cite{2005ExA....20..131H} is expected to launch in 2011, with a mirror with significant effective area up to 80~keV. This will focus the \Ti-decay lines at 68 and 78~keV, and could achieve sensitivity of $\simeq10^{-6}$~\flux or better in each. Objectives will be to map the \Ti in Cas~A, detect the lines in SN~1987A, and possibly in other young supernova remnants.  Japan's NeXT mission is expected to be launched a few years later and will include the HXT, with similar mirrors and capabilities. The Simbol-X mission~\cite{2008MmSAI..79...19F}, to be launched somewhat later, might extend mirrors to somewhat higher energies, and will still focus on \Ti lines from supernova remnants, among other objectives. 

\subsubsection{Gamma-Ray Lenses}
In visible light telescopes, another approach to increasing collection area is a lens. The same can be done at \gam-ray energies, using Bragg diffraction in high-Z crystals in a Laue lens~\cite{2006SPIE.6266E..60V}. The principle has been demonstrated with a balloon-borne instrument~\cite{2005ExA....20..253B}. A significant limitation is that a given lens crystal ring focuses only a narrow energy band on a detector at a fixed distance, but multiple rings can focus a small number of energy bands. The science objectives for such an instrument will have to be very focused, for example, on studying the \Ni and \Co lines near 800 keV, or the electron-positron annihilation line. A major such instrument has been proposed~\cite{2008ExA...tmp...44K}.

\subsubsection{Compton Telescopes}
A guiding principle for the different refinements of the original Compton Telescope principle~\cite{peterson1961,1973NucIM.107..385S} has been better characterization of all interactions, to provide improvement in celestial photon, and therefore background, recognition. After the demonstrated success of the COMPTEL experiment, efforts concentrate on refined detector modularity for improved event interaction locations, and on more advanced detector signal capturing and processing to enhance resolutions in time and energy. Especially useful for higher energies is tracking the recoil electrons, which limits possible photon directions. The MEGA prototype instrument of MPE \cite{2005NIMPA.541..310K} 
had been extensively calibrated and modelled through Monte Carlo simulations, and provides a perspective for reaching angular resolutions around 1 degree at a sensitivity of 5 and 8~10$^{-6}$~ph~cm$^{-2}$s$^{-1}$ for 10$^6$~s observing time at \Al and \Ti/\Fe line energies, respectively~\cite{2006NewAR..50..619B}. The GRIPS mission proposed for ESA's ``Cosmic Vision'' program features the currently most advanced design of a Compton telescope \cite{2008ExA...tmp...23G}. The ``Advanced Compton Telescope'' project proposed as a future NASA mission \cite{2006SPIE.6266E..62B} 
aims at even more ambitious sensitivities, near 10$^{-6}$~ph~cm$^{-2}$s$^{-1}$ for broad supernova lines, and including lower energies. This project has advanced instrument and background simulations to the point that they make credible predictions of on-orbit performance.

\section{Summary}
Gamma-ray lines from radioactive decay of cosmic nuclei have provided astrophysics with direct proof of ongoing nucleosynthesis in the present-day universe. Although the number of different isotopes available for such study is limited, these provide key calibration points for models of nucleosynthesis in stars and supernovae. \Ti is beginning to yield unique insights into the interior processes of massive-star gravitational collapses, with potential further insight into asymmetrical features such as jets, while corresponding measurements with Ni isotopes on thermonuclear supernova processes have not had the luck of a sufficiently nearby event. Diffuse \gam-ray emission in the Galaxy from annihilation of positrons, and from decay of radioactive \Al and \Fe all provide sensitive ensemble tests of our models of massive-star interior structure and nucleosynthesis (from \Al and \Fe), and propagation of positrons away from their nucleosynthesis sources within the complex interstellar-medium phases of the Galaxy. The interplay between such astronomical measurements and nuclear interaction cross sections, connected by astrophysical models, has been a rich area of research on nuclei in the cosmos. There is some concern that limitations of future space mission opportunities, with demands for big instruments on all frontiers of space astronomy, could lead to a ``dry period'' of decades, with corresponding instrumental expertise fading away quickly. Yet, nuclear astronomy clearly is an essential complement to atomic astronomy and astro-particle physics for our understanding of the cosmos.

\bibliography{../main,../diehl}
\bibliographystyle{JHEP}


\end{document}